# Universal Behavior in Large-scale Aggregation of Independent Noisy Observations

Tatsuto Murayama[*] and Peter Davis[†]

*NTT Communication Science Laboratories, NTT Corporation,*

*2-4, Hikaridai, Seika, Kyoto 619-0237, Japan*

(Dated: November 9, 2018)

## Abstract

Aggregation of noisy observations involves a difficult tradeoff between observation quality, which can be increased by increasing the number of observations, and aggregation quality which decreases if the number of observations is too large. We clarify this behavior for a protypical system in which arbitrarily large numbers of observations exceeding the system capacity can be aggregated using lossy data compression. We show the existence of a scaling relation between the collective error and the system capacity, and show that large scale lossy aggregation can outperform lossless aggregation above a critical level of observation noise. Further, we show that universal results for scaling and critical value of noise which are independent of system capacity can be obtained by considering asymptotic behavior when the system capacity increases toward infinity.

PACS numbers: 89.70.-a, 64.60.-i, 02.50.Cw, 75.10.Nr

---

[*]Electronic address: murayama@cslab.kecl.ntt.co.jp

[†]Electronic address: davis@cslab.kecl.ntt.co.jp

This letter presents results which give a new perspective on the growing field of sensory data aggregation by clarifying fundamental principles of large-scale aggregation. Examples of large scale aggregation of observations include astronomical observations [1], biological sensing [2], early detection of natural disasters such as earthquakes, tidal waves and floods [3] and wireless sensor networks [4]. Errors in observations can be reduced by collecting observation data from more sensors. However, collecting data from many sensors usually involves some cost in terms of network resources, resulting in fundamental tradeoffs [5]. The theoretical understanding of these tradeoffs in natural and engineered systems is now a high priority.

An important fundamental problem in this field is the problem of aggregating independent observations of the same phenomenon with a resource constraint. Previous works have analyzed the tradeoff behavior between aggregate data rate and sensing error from the fundamental view of information theory. The analysis has been extended to include the situation where arbitrarily large numbers of samples can be collected by reducing the data aggregated from each sample using lossy data compression. However, so far results have only been obtained for the fundamental information theoretic bounds with infinitely many sensors [6, 7], or specific situations in which the number of sensors is fixed [8]. The previous works do not include the situation where the number of observations can be varied, and thus the results are not sufficient to support our understanding and design of real world systems.

In this paper we introduce a modification of the common basic model for data aggregation with compression which makes it more tractable and amenable to analysis when the number of sensors can vary. Specifically, we consider independent decompression of each observation in a discrete version of the CEO problem [6]. We show that this model reveals a new property, the existence of noise threshold beyond which large scale aggregation is superior to lossless aggregation with no compression. This can be seen as a manifestation of "more is different" in sensor networks [9]. Moreover, we show that universal results for scaling behavior of collective estimation error can be obtained by considering asymptotic behavior when the system capacity diverges to infinity.

Suppose that we have $L$ independent sensors which each independently observe an $M$-bit state $X$, $X_\mu$ for $\mu = 1, \cdots, M$, of a common, uniform binary source, and obtain an $M$-bit observation $Y(a)$ ($a = 1, \cdots, L$) where each bit $Y_\mu(a)$ has common probability $p$ of error, i.e. differing from the corresponding source bit $X_\mu$. The value of $p$ specifies the level of observation noise. Now the sensors independently compress their $M$-bit observation into shorter $N$-bit codewords, $Z(a)$, and send them to the aggregator. The condition 'independent' excludes the possibility of mutual communications between sensors. We assume the rate $R = N/M$ is common to all the sensors. In addition, we suppose that the sum total of the rate, the system capacity $\lambda$, is fixed, with

$$\lambda = LR .\qquad(1)$$

The aggregator then decodes every $N$ bit codeword independently to obtain $L$ separate $M$-bit reproductions $\hat{Y}(a)$ ($a = 1, \cdots, L$). Finally, the $\hat{Y}(a)$ are used to obtain a single collective estimator $\hat{X}$. We analyze the behavior of the bit error probability, denoted $p_e(p, R; \lambda)$, in the collective estimate.

The theoretical lower bound of average distortion for a given rate $R$ is given by the distortion-rate function, or simply the Shannon bound [10]. Though we know that the bound could be achieved asymptotically by using Shannon's random codes, the exponential encoding complexity prohibits us from using them in practice. For uniform binary sources, however, an alternative approach has been recently developed based on linear codes with iterative, or message passing, encoding achieving close to the theoretical limit [11, 12, 13]. Applying these new results allows us to obtain numerical results for arbitrary data reduction.

FIG. 1 shows typical results from a numerical experiment for the average values of per-bit error probability, $p_e(p, R; \lambda)$, obtained using a linear code with an iterative encoder [11]. The linear codes are defined by a class of sparse matrices having $K$ ones ('1') per row and $C$ ones per column, respectively, where $K/C = N/M$. Therefore we may write $R = K/C$ [14]. For ease of comparison, the values of error probability $p_e(p, R; \lambda)$ for noise $p$ and rate $R = N/M$ are divided by a reference level $p_e(p, 1; \lambda)$ for $R = 1$ under the same system capacity $\lambda$ [15].

The example FIG. 1 demonstrates the following two points; (1) *There exists a threshold value of noise where lossy large-scale aggregation becomes superior to lossless aggregation.* Lossless aggregation with $R = 1$ outperforms the lossy aggregation with $R$ smaller than 1 at lower noise levels. However, at higher noise levels the alternative strategy with lossy data compression becomes superior. (2) *There exists a scaling relation with respect to system capacity.* The error curves have a universal shape in the sense that plots for different $\lambda$ overlap with appropriate re-scaling, as shown by the example for $\lambda = 500$ using the scale on the right side. This observation implies a scaling law for the data aggregation with respect

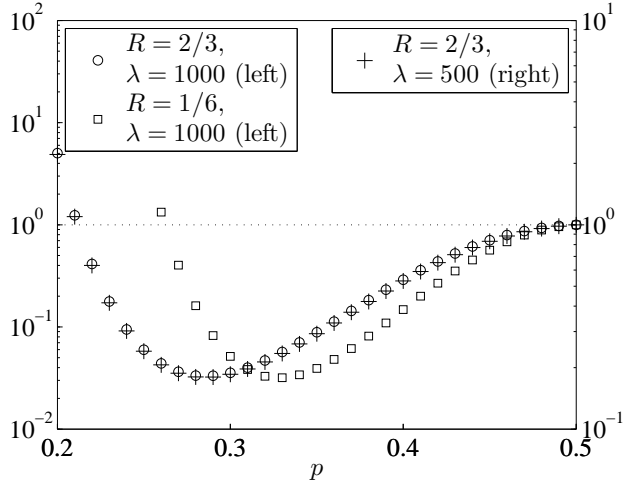

FIG. 1: Semilog plots for average error probability in noisy data aggregation using linear codes with $K = 2$. The values of error probability $p_e(p, R; \lambda)$ for noise $p$ and rate $R = K/C$ are divided by a reference level $p_e(p, 1; \lambda)$ under the same system capacity $\lambda$. Here parameters are chosen to be $\lambda = 500$ with $C = 3$ (pluses, right scale) and $\lambda = 1000$ with $C = 3$ and $12$ (circles and squares, left scale), respectively.

to $\lambda$. Introducing the coefficient $\beta$ we can write the empirical scaling relation as follows:

$$\log\left[\frac{p_e(p, R; \beta\lambda)}{p_e(p, 1; \beta\lambda)}\right] = \beta \log\left[\frac{p_e(p, R; \lambda)}{p_e(p, 1; \lambda)}\right] . \tag{2}$$

Using the base-10 logarithm, the scaling in FIG. 1 is well defined by the scaling factor $\beta = 2$.

In this letter, we present a theoretical analysis which explains these empirical results, and presents them in a universal form. First, we assume that the error due to lossy compression is independent of $\mu$ and $a$, and denoted by $D$, that is $\langle \delta(Y_\mu(a), -\hat{Y}_\mu(a)) \rangle = D$. Here we used Kronecker's delta $\delta$ and the braket denotes averaging over random variables. This includes the standard exchangeable sensor ansatz for our model [6, 7], which means that all sensors have the same rate $R$ and distortion $D$. The possible value of the distortion $D$ depends on $R$, so we explicitly denote $D$ as $D(R)$. The combined error probability for $\hat{Y}_\mu(a)$, independent of $\mu$ and $a$, is obtained as

$$\rho = (1 - 2p)D(R) + p . \tag{3}$$

The combined error probability $\rho$ is a function of both $p$ and $R$. In particular, the equation (3) implies that $\rho$ is a decreasing function of $R$, since $D(R)$ should be a decreasing function of $R$.

Since we assume Bernoulli statistics, the best estimate from the set of aggregated values can be obtained by the simple majority-vote operation:

$$\hat{X}_\mu = \text{sgn}\left\{\sum_{a=1}^{L} \hat{Y}_\mu(a)\right\}.$$

Then, the error probability for the final estimate is given, in terms of $\rho$ and $L$, by $p_e(p, R; \lambda) = \sum_{l=\frac{L+1}{2}}^{L} Q_\rho(l; L)$, which is just the probability of getting more than $L/2$ errors out of $L$ Bernoulli trials. We assume for simplicity that only odd values of $L$ are taken. The $Q_\rho(l; L) = \binom{L}{l}\rho^l(1-\rho)^{L-l}$ represents the binomial distribution.

It is obvious that $p_e(p, R; \lambda)$ is a decreasing function of $L$ if $\rho$ is fixed. However, due to the constraint (1), and the decrease in distortion $D(R)$ with increase of $R$, $\rho$ actually increases with an increase of $L$, resulting in contrary effects on $p_e(p, R; \lambda)$. Therefore the challenge here is to incorporate consideration of the distortion $D$ in a way which clarifies the interplay between the contrary effects induced by the constraint (1).

In the following, we consider the asymptotic analysis in the limit of large $\lambda$, for which we can obtain explicit results. For sufficiently large $L$, the binomial distribution $Q_\rho(l|L)$ is well approximated by the Gaussian distribution $N(L\rho, L\rho(1-\rho))$ with mean $L\rho$ and variance $L\rho(1-\rho)$. Now we examine the asymptotic behavior for $\lambda$. Write $\alpha(p, R) = (1-2p)(1-2D(R))$ and define, for simplicity,

$$\nu = \frac{\alpha(p, R)\sqrt{\lambda}}{\sqrt{R(1-\alpha(p,R))(1+\alpha(p,R))}}.$$

Then, in the limit $\lambda \to \infty$, the asymptotic expansion of the cummulative Gaussian distribution gives

$$p_e(p, L; \lambda) \sim \frac{1}{2}\text{erfc}\left(\frac{\nu}{\sqrt{2}}\right),$$

where $\text{erfc}(x)$ is the complimentary error function [16]. By analogy with large deviation theory [17], we can define and calculate the exponential rate of decay as follows:

$$I_p(R) = -\lim_{\lambda \to \infty} \frac{1}{\lambda} \ln p_e(p, R; \lambda)$$
$$= \frac{\alpha(p, R)^2}{2R(1-\alpha(p,R))(1+\alpha(p,R))} \quad (0 < R \leq 1). \quad (4)$$

Notice that the above formula holds for any function $D(R)$. Indeed this universal property well describes the exponential scaling (2).

In particular, the smallest average distortion $D(R)$ is obtained in the limit of $M \to \infty$, and is called the distortion-rate function [10]. In our model, its inverse function, the rate-distortion function [10], can be analytically given by

$$R(D) = 1 + D \log_2 D + (1 - D) \log_2(1 - D) . \tag{5}$$

We may use either the distortion-rate function or the rate-distortion function to describe the optimal boundary, since the two descriptions are equivalent in the large $M$ limit.

Now assume hereafter that the distortion-rate function $D(R)$ is the specific case implicitly given by the inverse formula (5) for $R(D)$. Then asymptotics of $R(D)$ enables us to obtain the large scale decay rate as

$$I_p(0) = -\lim_{\lambda \to \infty} \frac{1}{\lambda} \lim_{R \to 0} \ln p_e(p, R; \lambda) = (1 - 2p)^2 \ln 2 .$$

Now we can see that if we compare just the two aggregation strategies $R = 1$ or $R \to 0$, the threshold value of noise $p_1$ corresponding to the switch of the superior aggregation can be determined by solving the equation

$$(1 - 2p_1)^2 \ln 2 = \frac{(1 - 2p_1)^2}{2(1 - (1 - 2p_1)^2)} .$$

The analytical solution $p_1 = 0.236$ gives the threshold beyond which the large scale aggregation with $R \to 0$ outperforms the $R = 1$ strategy.

Next we numerically examine the value of $R$ which maximizes $I_p(R)$ for a given $p$. The optimal value $R^*$ is plotted in FIG. 2 as a function of $p$. We find that the optimal rate vanishes, i.e. $R^* = 0$, for noise levels larger than a critical point $p_0 = 0.295$. In contrast, we can always find non-zero optimal values of $R$ below this point. In particular, if the noise level is near zero, then $R = 1$ is optimal. The change in value of optimal $R^*$ with respect to noise level $p$ is continuous at $p_0$, as in a second order phase transition.

We note that the analytical results presented here using (4) and (5) are consistent with the results of the numerical simulations with linear codes. That is, the exponential rate of decay (4) well describes the scaling law (2). Moreover, they add more specific and fundamental conditions to our first observation on FIG. 1 that aggregation with $R$ smaller than 1 is superior for larger noise. The critical point beyond which the strategy with $R = 1$ is not optimal in FIG. 2 indicates the lowest bound for such threshold, and is obviously consistent with the numerical simulations.

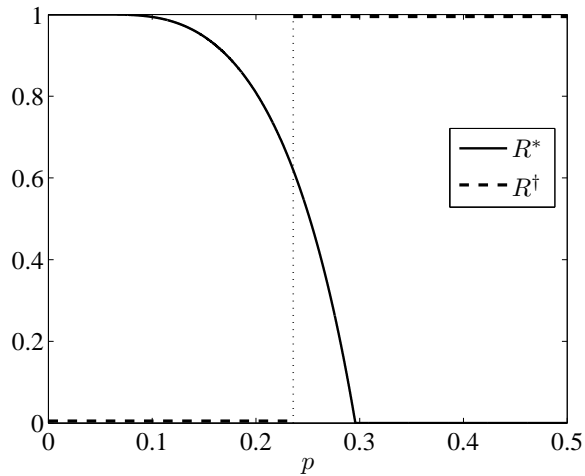

FIG. 2: Optimal rate $R^*$ for lossy aggregation of observations from independent sensors in noisy environment with noise level $p$. $R^*$ is the optimal value of $R$, the aggregation rate per sensor, which maximizes the asymptotic decay rate $I_p(R)$ of error probability with increase of system capacity $\lambda$. $I_p(R)$ is defined in (4). Distortion due to lossy compression is given implicitly by (5). For comparison, $R^\dagger$ is the pessimistic value of $R$ which minimizes $I_p(R)$.

Now let us consider the value of $R$ which minimizes $I_p(R)$, say $R^\dagger$. In contrast with the continuous change in the behavior of optimal $R^*$, the pessimistic $R^\dagger$ shows an abrupt change with respect to the noise $p$. Our numerical analysis indicates that there are only two cases for the worst solution: $R^\dagger = 0$ and $R^\dagger = 1$, so the threshold value of noise $p_1$ corresponds to the switch of the $R^\dagger$.

We note that in the intermediate range of $p$ the optimal $R^*$ is a finite value between $R = 1$ and $R = 0$. It is natural to ask how much the estimates obtained with these intermediate values of $R^*$ differ from the estimates obtained using the extreme values of $R = 1$ or $R = 0$. FIG. 3 shows the noise dependence of decay rates $I_p(R)$ with $R = 0$, 1, and $R^*$, respectively. The size of the difference $I_p(R^*) - I_p(1)$ and $I_p(0) - I_p(1)$ is shown in the inset of FIG. 3. For comparison with these results which were obtained using the Shannon limit, the rate-distortion function in (5), we also show the result obtained for linear code with $K = 2$, corresponding to FIG. 1. This result for $K = 2$ was obtained using the replica method for diluted spin systems [18, 19]. First we note that in the case of compression using $R(D)$, expression (5), the combination strategy of using only either $R = 0$ or $R \to 0$, switching at the threshold point $p_1$, well approximates the optimal performance given by $R^*$. Next, we focus on the behavior of the difference $I_p(R^*) - I_p(1)$ with respect to the noise $p$ (solid line in inset). The largest gain is achieved at $p^* = 0.305$ (indicated in the figure by a vertical dotted line), which differs slightly from the value for $p_0$ which was $p_0 = 0.295$. Finally, we consider the result for the linear code with $K = 2$. It shows a similar threshold behavior - the value of $I_p(R)$ for $R = 0$ becomes greater than the value for $R = 1$ when the noise $p$ exceeds a threshold value [20]. However, the gain is less than that obtained for the rate-distortion function, which shows that there is still room for improvement by using alternative techniques [21, 22].

Our results show that the optimal aggregation for a system of sensors with constrained system capacity exhibits a kind of threshold behavior with respect to the observation noise level. If we imagine the system autonomously switching to the optimal aggregation method, then it would appear to be a phase transition behavior. This result is significant for understanding the principles of large scale aggregation in sensing systems, natural or engineered. We described the behavior of the optimal aggregation rate per sensor $R = \lambda/L$, the ratio of the system capacity $\lambda$ and the number of sensors $L$. The analysis shows that in the high noise region beyond a critical value of noise, the rate $R$ should approach to zero in order to reduce collective estimation error. This means that very many sensors with $L \gg \lambda$ should be used. In contrast, if the noise level is lower than the critical point, the ratio $R$ should take a positive value. In this case, the number of sensors scales as $L = O(\lambda)$.

This work has been supported in part by Grant-in-Aid for Scientific Research on Priority Areas, Ministry of Education, Culture, Sports, Science and Technology (MEXT), Japan, No. 18079015.

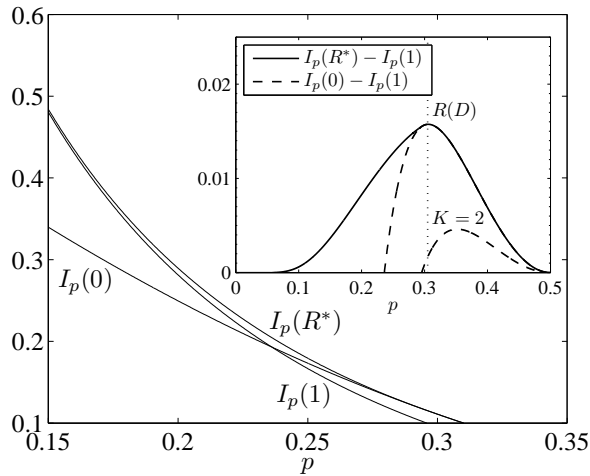

FIG. 3: Error decay rates. $I_p(R^*)$ is the error decay rate with lossy data compression at the optimal aggregation rate $R^*$, while $I_p(1)$ is the error decay rate without data compression. $I(0)$ corresponds to the error decay rate for the large system limit when $R \to 0$. Inset: Information gain. $R(D)$ corresponds to the Shannon limit, while $K = 2$ indicates performance of the linear codes when $C \to \infty$.